\pdfoutput=1
\documentclass[conference,a4paper,10pt]{IEEEtran}

\usepackage[latin1]{inputenc}
\usepackage{times,amsmath}
\usepackage{pstool}
\usepackage{subfigure}
\usepackage{multirow}
\usepackage{enumerate}
\usepackage{graphicx}
\usepackage{MnSymbol}
\usepackage{stfloats} 
\usepackage[table]{xcolor}
\usepackage[square, comma, sort&compress, numbers]{natbib}
\usepackage{nohyperref}
\usepackage{algorithm,algorithmic}
\usepackage{mathtools}

\newcounter{tempEquationCounter} 
\newcounter{thisEquationNumber}

\makeatletter
\newcommand{\vast}{\bBigg@{4}}
\newcommand{\Vast}{\bBigg@{5}}
\makeatother

\begin{document}

\title{ Distributed Beamforming with Wirelessly Powered Relay Nodes}

\author{
\IEEEauthorblockN{
Muhammad Ozair Iqbal\IEEEauthorrefmark{1},
Ammar Mahmood\IEEEauthorrefmark{1}, 
Muhammad Mahboob Ur Rahman\IEEEauthorrefmark{1},
Qammer H. Abbasi\IEEEauthorrefmark{2}
} 
\IEEEauthorblockA{
\IEEEauthorrefmark{1}Electrical Engineering department, Information Technology University (ITU), Lahore, Pakistan \\\ \{ozair.iqbal,ammar.mahmood,mahboob.rahman\}@itu.edu.pk }
\authorblockA{
\IEEEauthorrefmark{2}Department of Electrical and Computer Engineering, Texas A\&M University at Qatar \\\ qammer.abbasi@tamu.edu }
}

\maketitle

\begin{abstract} 

This paper studies a system where a set of $N$ relay nodes harvest energy from the signal received from a source to later utilize it when forwarding the source's data to a destination node via distributed beamforming. To this end, we derive (approximate) analytical expressions for the mean SNR at destination node when relays employ: i) time-switching based energy harvesting policy, ii) power-splitting based energy harvesting policy. The obtained results facilitate the study of the interplay between the energy harvesting parameters and the synchronization error, and their combined impact on mean SNR. Simulation results indicate that i) the derived approximate expressions are very accurate even for small $N$ (e.g., $N=15$), ii) time-switching policy by the relays outperforms power-splitting policy by at least $3$ dB.

\end{abstract}

\section{Introduction}
\label{sec:intro}

Distributed transmit beamforming is a technique whereby multiple transmitters cooperate in a way that their signals (carrying a common message) combine coherently, over-the-air, at the intended receiver. For the unit-gain channels between the transmit nodes and the receiver, distributed beamforming leads to an $N^2$-fold increase in mean SNR at the receiver (where $N$ is the number of cooperating transmitters) \cite{Raghu:TWC:2007}. However, the energy-efficiency advantage of distributed beamforming comes at a cost, the carrier synchronization cost. Specifically, the individual passband signals sent from cooperating transmit nodes combine constructively at the receiver only when transmit nodes are frequency, time and phase synchronized \cite{Raghu:TWC:2007}.

Quite recently, wireless power transfer where a transmit node lets its receive counterpart harvest energy from the radio frequency (RF) signal it transmits, has attracted a lot of attention \cite{Ali:TWC:2013}. In the literature, two energy harvesting scenarios have been widely studied: i) time-switching (TS) based energy harvesting (EH) where the receiver spends a (time) fraction of every symbol it receives for energy harvesting, ii) power-splitting (PS) based energy harvesting where the receiver spends a fraction of the received power for energy harvesting.

This paper studies a system where a set of $N$ relay nodes harvest energy from the signal received from a source to later utilize it when forwarding the source's data to a destination node via distributed beamforming. Specifically, the paper derives (approximate) analytical expressions for the mean SNR at destination node when relays employ: i) TS based EH scheme, ii) PS based EH scheme. The obtained results facilitate us to study the interplay between the energy harvesting parameters and the synchronization error, and their influence on mean SNR. Simulation results indicate that the derived approximate expressions are very accurate even for small $N$ (e.g., $N=15$).

The related works closest to this work are \cite{Suraweera:TComm:2015},\cite{Brown:CISS:2015}. \cite{Suraweera:TComm:2015} considers a single multi-antenna relay which harvests energy from a source (and external interferences) to later forward its data (via maximum ratio transmission) to the destination; authors of \cite{Suraweera:TComm:2015} then derive closed-form expressions for outage probability and ergodic capacity of the system. In \cite{Brown:CISS:2015}, multiple transmit nodes do (received-assisted) distributed beamforming towards a receiver node where the receiver node harvests energy from the received sum signal; \cite{Brown:CISS:2015} then studies the trade-off between feedback rate and amount of energy harvested at the receiver. Nevertheless, to the best of authors' knowledge, the interplay between energy harvesting parameters and synchronization error, and their collective impact on mean SNR (presented in this work) has not been studied before.

\section{System model}
\label{sec:sys-model}

\begin{figure}[h]
\centering{\includegraphics[width=90mm]{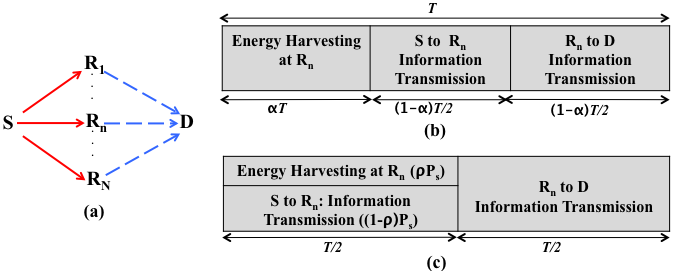}}
\caption{System model}
\label{fig:sys-model}
\end{figure}

A system consisting of a source node $S$, a destination node $D$ and $N$ relay nodes ($R_1,...,R_N$) is studied (see Fig. \ref{fig:sys-model}(a)). Following assumptions are in place: direct link between $S$ and $D$ is not available; the relay nodes operate in half-duplex mode and employ decode-and-forward (DF) strategy; the relays do distributed beamforming towards $D$; the relays are fully, wirelessly powered by the $S$; the channels on both hops are quasi-static (i.e., each channel stays constant for a slot duration $T$ and channel realizations are $i.i.d$ between the slots), frequency-flat, block fading with Rayleigh distribution. 
\section{Time switching based energy harvesting at relays}
\label{sec:TSR}

Let $T$ denote the block time during which source $S$ transmits a certain amount of information to destination $D$. Then, under time-switching (TS) based energy harvesting (EH) policy, the relays harvest energy from source's transmission for a duration $\alpha T$, where $0 \le \alpha \le 1$ (see Fig. \ref{fig:sys-model}(b)). 

Specifically, on the first hop, source $S$ transmits message $x$ (with power $P_s$) to the relays. Then, relay $R_n$ ($n=1,...,N$) receives $y_n = \sqrt{P_S}{g_n} x + w_n$, where $g_n$ is the channel between source and relay $R_n$, and $w_n$ is the noise at relay $R_n$.  Then, the amount of energy harvested by $R_n$ is $EH_n = \eta |\sqrt{P_S}{g_n}|^2\times \alpha T$ where $0 \le \eta \le 1$ is the (RF to DC) energy conversion efficiency. Since the relay $R_n$ uses all of the energy harvested to relay the (perfectly recovered) message $x$  to the destination, the transmit power of $R_n$ is $P_n = \frac{EH_n}{(1-\alpha)T/2} = \frac{2\eta |\sqrt{P_S}g_n|^2\alpha}{(1-\alpha)}$.
Next, on the second hop, each of the $N$ relays {\it simultaneously} forwards the precoded message $r_n=a_n x$ to the destination $D$ ($a_n$ is the precoding weight applied by $R_n$). The net (sum) signal received at $D$ is:
\begin{equation}
	\label{eq:z}
	z = \sum_{n=1}^N \sqrt{P_n}h_n r_n + w_D 
\end{equation}
where $h_n=|h_n|e^{j\phi_n}$ is the channel between the relay $R_n$ and destination $D$, and $w_D$ is the noise at $D$; $w_D\sim \mathcal{CN}(0, \sigma_D^2)$. Let $g_n, h_n \sim \mathcal{CN}(0,1)$, $\forall$ $n=1,...,N$. Then, one can verify that $P_n\sim \exp{(\lambda_p)}$ where $\lambda_p = \frac{1-\alpha}{2\eta \alpha P_S}$, and $\sqrt{P_n}\sim \textrm{Rayleigh}(\sigma)$ where $\sigma = \sqrt{\frac{\eta \alpha P_S}{(1-\alpha)}}$. 
When relays do distributed beamforming, relay $R_n$ chooses $a_n \triangleq \frac{1}{|h_n|}e^{-j(\phi_n-\theta_n)}$ (this could be achieved by running the protocols proposed in, e.g., \cite{Raghu:TWC:2007},\cite{Quitin:TWC:2013},\cite{Quitin:TWC:2016}). Then, Eq. (\ref{eq:z}) can be rewritten as:
\begin{equation}
	\label{eq:z2}
	z = \sum_{n=1}^N \sqrt{P_n}e^{j\theta_n}x + w_D 
\end{equation}
where $\theta_n$ models the channel phase estimation error for the channel $h_n$. However, we note that $\theta_n$ could very well represent the net phase difference between $R_n$ and $D$, i.e., it could assimilate the effects of channel phase estimation error, frequency and phase offsets etc.). Indeed, in this work, we assume that $\theta_n$ denotes the effective phase difference between $R_n$ and $D$. Moreover, we assume that $\theta_n$ are $i.i.d$ with $\sim \mathcal{N}(0,\sigma_{\theta}^2)$. Next, assuming that $x \in$ $M$-PSK constellation (for any $M$) and that $\sigma_D^2=1$, the instantaneous SNR at the destination $D$ is:
\begin{equation}
	\label{eq:gamma}
	\gamma_D(\{\theta_n\},\{P_n\}) = \gamma_D(\{\theta_n\},\alpha) = {\bigg| \sum_{n=1}^{N} \sqrt{P_n}e^{j\theta_n}\bigg|^2}
\end{equation}
Then, an (approximate) expression for mean SNR $E[\gamma_D]$ is provided in Corollary 1.

{\it Corollary 1:} Let $\hat{\gamma}_D \triangleq \lim_{N \to \infty} \gamma_D$. Then, the following holds:
\begin{equation}
	\label{eq:ESNR1}
	E[\hat{\gamma}_D(\{\theta_n\},\alpha)] = a^2 + b^2 + c + d
\end{equation}
where
\begin{align*}
\label{eq:meanIQ}
a = N \sqrt{\frac{\pi \eta \alpha P_S}{2(1-\alpha)}} e^{-\sigma_{\theta}^2/2}; b = 0;
\end{align*}
\begin{equation*}
	\label{eq:VarI}	
	c  = N\bigg[ \frac{\eta \alpha P_S}{(1-\alpha)}(1-e^{-\sigma_{\theta}^2})^2 + \frac{2 \eta \alpha P_S}{(1-\alpha)} e^{-\sigma_{\theta}^2/2} \bigg]-N\bigg(\sqrt{\frac{\pi \eta \alpha P_S}{2(1-\alpha)}}.e^{-\sigma_{\theta}^2/2} \bigg)^2 ; 
\end{equation*} 
\begin{align*}
	\label{eq:VarQ}
	d = \frac{N \eta \alpha P_S}{(1-\alpha)}(1-e^{-2\sigma_{\theta}^2}) .
\end{align*}
{\it Proof:} See Appendix A.
\section{Power splitting based energy harvesting at relays}
\label{sec:PSR}

Under power-splitting based energy harvesting policy, relay $R_n$ harvests $EH_n = \eta \rho |\sqrt{P_S}{g_n}|^2\times T/2$ amount of energy from source's transmission for a duration $T/2$ (see Fig. \ref{fig:sys-model}(c)). Since the relay $R_n$ uses all of the energy harvested to relay the message $x$  to $D$, the transmit power of $R_n$ is $P_n = \frac{EH_n}{T/2} = 2\eta \rho |\sqrt{P_S}g_n|^2$. In this case, $P_n\sim \exp{(\lambda_p)}$ where $\lambda_p = \frac{1}{2\eta \rho P_S}$, and $\sqrt{P_n}\sim \textrm{Rayleigh}(\sigma)$ where $\sigma = \sqrt{\eta \rho P_S}$. Then, one can verify that the sum signal $z$ received at $D$ and instantaneous SNR $\gamma_D$ are once again given by Eq. (\ref{eq:z2}) and Eq. (\ref{eq:gamma}) respectively. Then, an (approximate) expression for mean SNR $E[\gamma_D]$ is provided in Corollary 2. 

{\it Corollary 2:} Let $\hat{\gamma}_D \triangleq \lim_{N \to \infty} \gamma_D$. Then, the following holds:
\begin{equation}
	\label{eq:ESNR2}
	E[\hat{\gamma}_D(\{\theta_n\},\rho)] = p^2 + q^2 + r + s
\end{equation}
where
\begin{equation*}
p = N \sqrt{\frac{\pi \eta \rho P_S}{2}} e^{-\sigma_{\theta}^2/2}; q = 0;
\end{equation*}
\begin{equation*}
	\label{eq:VarI}
	r = N\bigg[ \eta \rho P_S (1-e^{-\sigma_{\theta}^2})^2 + 2 \eta \rho P_S e^{-\sigma_{\theta}^2/2} \bigg]-N\bigg(\sqrt{\frac{\pi \eta \rho P_S}{2}}.e^{-\sigma_{\theta}^2/2} \bigg)^2 ;
\end{equation*} 
\begin{equation*}
	\label{eq:VarQ}
	s = N \eta \rho P_S (1-e^{-2\sigma_{\theta}^2}) .
\end{equation*} 
{\it Proof:} See Appendix A.

\section{Simulation results}
\label{sec:sim}

\begin{figure}[ht]
\begin{center}
	\includegraphics[width=3.5in,height=2.2in]{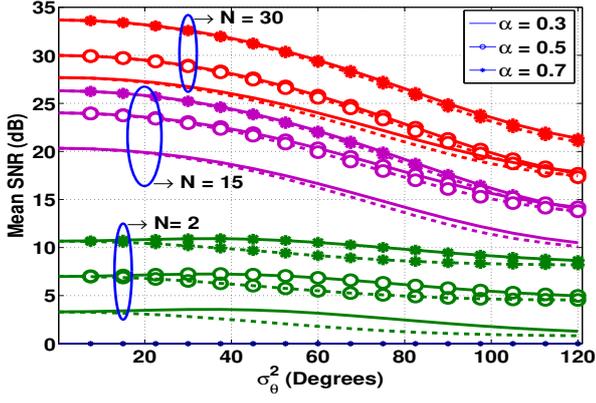} 
\caption{TS based EH scheme.}
\label{fig:snrvsphase}
\end{center}
\end{figure}

\begin{figure}[ht]
\begin{center}
	\includegraphics[width=3.5in,height=2.2in]{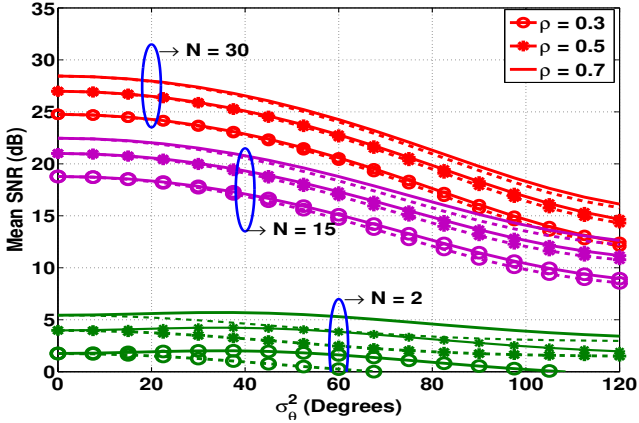} 
\caption{PS based EH scheme.}
\label{fig:snrvsphase2}
\end{center}
\end{figure}

In all plots, solid lines represent analytical predictions by Eqs. (\ref{eq:ESNR1}), (\ref{eq:ESNR2}) while dotted lines represent Monte-Carlo simulation results. Figs. \ref{fig:snrvsphase}, \ref{fig:snrvsphase2} show the following: i) the analytical approximations of Eqs. (\ref{eq:ESNR1}), (\ref{eq:ESNR2}) are indeed very accurate for $N$ as low as $15$ (while the approximations degrade for $N=2$); ii) the mean SNR degrades as the variance of the net phase error increases (due to poorer oscillators, poor synchronization protocol etc.); iii) for a given system state (of phase error variance), the mean SNR can be improved by doing more energy harvesting at the relays; iv) the TS based EH scheme outperforms PS based EH scheme by at least $3$ dB, for a given phase error variance.

\section{Conclusion}
\label{sec:conc}

This preliminary work studied a system where a set of $N$ relay nodes harvest energy from the signal received from a source to later utilize it when forwarding the source's data to a destination node via distributed beamforming. Monte-Carlo simulation results showed that the derived approximate expressions for the mean SNR at the destination are very accurate for $N$ as low as $15$. Last but not the least, TS based EH scheme outperformed PS based EH scheme by at least $3$ dB. Immediate future work will investigate the coupling (dependence) between energy harvesting parameters and the phase error (due to clock drift) and their combined impact on mean SNR (and Ergodic capacity) at the destination.

\appendices
\section{An Approximate Expression for the mean SNR}
\label{app:clt}
We can rewrite $\gamma_D$ from Eq. (\ref{eq:gamma}) as:
\begin{equation}
	\label{eq:gamma_re_im}
	\gamma_D = \bigg( \sum_{n=1}^{N} \sqrt{P_n} \cos{\theta_n} \bigg)^2 + \bigg( \sum_{n=1}^{N} \sqrt{P_n} \sin{\theta_n} \bigg)^2
\end{equation} 
Let $X_n=\sqrt{P_n} \cos{\theta_n}$ and $Y_n=\sqrt{P_n} \sin{\theta_n}$. Then, $\gamma_D = (\sum_{n=1}^{N} X_n)^2 + (\sum_{n=1}^{N} Y_n)^2$. Note that even though $\sqrt{P_n}\sim \textrm{Rayleigh}(\sigma)$ and $\theta_n\sim \mathcal{N}(0, \sigma_{\theta}^2)$, the distribution of each of $X_n$ and $Y_n$ is not easy to obtain. However, note that both $\sqrt{P_n}$ and $\theta_n$ are {\it i.i.d}; therefore, if one knows the means $E[X_n]$,$E[Y_n]$ and variances $Var[X_n]$,$Var[Y_n]$ of $X_n$ and $Y_n$ respectively, then (for large $N$) one can invoke Central Limit Theorem to get a step closer towards obtaining expected value of $\gamma_D$. To this end, we have: 
\begin{equation}
	\label{eq:EXn}
	E[X_n] = E[\sqrt{P_n} \cos{\theta_n}] = E[\sqrt{P_n}].E[\cos{\theta_n}]=\sigma.\sqrt{\pi/2}.e^{-\sigma_{\theta}^2/2}
\end{equation}
where we have used the fact that $\sqrt{P_n}$ and $\cos{\theta_n}$ are independent of each other. And
\begin{equation}
	\label{eq:EYn}
	E[Y_n] = E[\sqrt{P_n} \sin{\theta_n}] = E[\sqrt{P_n}].E[\sin{\theta_n}] = 0
\end{equation} 
Similarly, we have:
\begin{equation}
	\label{eq:VarXn}
\begin{split}
	Var[X_n] & = E[X_n^2]-(E[X_n])^2 \\
	& = \frac{1}{\lambda_p}(\frac{1}{2}(1-e^{\sigma_{\theta}^2})^2+e^{-\sigma_{\theta}^2/2}) - (\sigma.\sqrt{\frac{\pi}{2}}.e^{-\sigma_{\theta}^2/2})^2 
\end{split}
\end{equation} 
\begin{equation}
	\label{eq:VarYn}
	Var[Y_n]  = E[Y_n^2]-(E[Y_n])^2 = \frac{1}{2\lambda_p}(1-e^{-2\sigma_{\theta}^2})
\end{equation} 
Let $I = \sum_{n=1}^{N} X_n$ and $Q = \sum_{n=1}^{N} Y_n$. Then, $\gamma_D=I^2+Q^2$. Let $\hat{\gamma}_D \triangleq \lim_{N \to \infty} \gamma_D$. Then, according to Central Limit Theorem, the following relations hold: 
$\lim_{N \to \infty} I\sim \mathcal{N}(m_I,\sigma_I^2)$; $\lim_{N \to \infty} Q\sim \mathcal{N}(m_Q,\sigma_Q^2)$ 
where
$m_I = N E[X_n] = N \sigma \sqrt{\frac{\pi}{2}} e^{-\sigma_{\theta}^2/2}; m_Q = N E[Y_n] = 0;$
\begin{equation}
	\label{eq:VarI}
\begin{split}
	\sigma_I^2 & = N. Var[X_n] \\
	& = N\bigg[ \frac{1}{2\lambda_p}(1-e^{-\sigma_{\theta}^2})^2 + \frac{1}{\lambda_p} e^{-\sigma_{\theta}^2/2} \bigg] - N\bigg(\sigma.\sqrt{\frac{\pi}{2}}.e^{-\sigma_{\theta}^2/2} \bigg)^2;
\end{split}
\end{equation} 
\begin{equation}
	\label{eq:VarQ}
	\sigma_Q^2 = N. Var[Y_n] = \frac{N}{2\lambda_p}(1-e^{-2\sigma_{\theta}^2}).
\end{equation} 
Then, 
\begin{equation}
	\label{eq:ESNR}
	E[\hat{\gamma}_D] = E\bigg[ \lim_{N \to \infty}I^2 \bigg] + E\bigg[ \lim_{N \to \infty}Q^2 \bigg] = \sigma_I^2 + m_I^2 + \sigma_Q^2 + m_Q^2
\end{equation}

\section*{Acknowledgements}
This publication was made possible by NPRP grant $\# 7-125-2-061$ from the Qatar National Research Fund (a member of Qatar Foundation). The statements made herein are solely the responsibility of the authors.

\footnotesize{
\bibliographystyle{IEEEtran}
\bibliography{refs}

\begin{thebibliography}{1}
\providecommand{\url}[1]{#1}
\csname url@rmstyle\endcsname
\providecommand{\newblock}{\relax}
\providecommand{\bibinfo}[2]{#2}
\providecommand\BIBentrySTDinterwordspacing{\spaceskip=0pt\relax}
\providecommand\BIBentryALTinterwordstretchfactor{4}
\providecommand\BIBentryALTinterwordspacing{\spaceskip=\fontdimen2\font plus
\BIBentryALTinterwordstretchfactor\fontdimen3\font minus
  \fontdimen4\font\relax}
\providecommand\BIBforeignlanguage[2]{{%
\expandafter\ifx\csname l@#1\endcsname\relax
\typeout{** WARNING: IEEEtran.bst: No hyphenation pattern has been}%
\typeout{** loaded for the language `#1'. Using the pattern for}%
\typeout{** the default language instead.}%
\else
\language=\csname l@#1\endcsname
\fi
#2}}

\bibitem{Raghu:TWC:2007}
R.~Mudumbai, G.~Barriac, and U.~Madhow, ``On the feasibility of distributed
  beamforming in wireless networks,'' \emph{Wireless Communications, IEEE
  Transactions on}, vol.~6, no.~5, pp. 1754--1763, May 2007.

\bibitem{Ali:TWC:2013}
A.~A. Nasir, X.~Zhou, S.~Durrani, and R.~A. Kennedy, ``Relaying protocols for
  wireless energy harvesting and information processing,'' \emph{IEEE
  Transactions on Wireless Communications}, vol.~12, no.~7, pp. 3622--3636,
  July 2013.

\bibitem{Suraweera:TComm:2015}
G.~Zhu, C.~Zhong, H.~A. Suraweera, G.~K. Karagiannidis, Z.~Zhang, and T.~A.
  Tsiftsis, ``Wireless information and power transfer in relay systems with
  multiple antennas and interference,'' \emph{IEEE Transactions on
  Communications}, vol.~63, no.~4, pp. 1400--1418, April 2015.

\bibitem{Brown:CISS:2015}
R.~Wang, R.~David, and D.~R. Brown, ``Feedback rate optimization in
  receiver-coordinated distributed transmit beamforming for wireless power
  transfer,'' in \emph{2015 49th Annual Conference on Information Sciences and
  Systems (CISS)}, March 2015, pp. 1--6.

\bibitem{Quitin:TWC:2013}
F.~Quitin, M.~M.~U. Rahman, R.~Mudumbai, and U.~Madhow, ``A scalable
  architecture for distributed transmit beamforming with commodity radios:
  Design and proof of concept,'' \emph{IEEE Transactions on Wireless
  Communications}, vol.~12, no.~3, pp. 1418--1428, 2013.

\bibitem{Quitin:TWC:2016}
F.~Quitin, A.~T. Irish, and U.~Madhow, ``A scalable architecture for
  distributed receive beamforming: Analysis and experimental demonstration,''
  \emph{IEEE Transactions on Wireless Communications}, vol.~15, no.~3, pp.
  2039--2053, March 2016.

\end{thebibliography}
}

\vfill\break

\end{document}